**An exponential–logarithmic measure of drug–receptor binding and saturation**

Arturo Tozzi (corresponding author)
ASL Napoli 1 Centro, Distretto 27, Naples, Italy
Via Comunale del Principe 13/a 80145
tozziarturo@libero.it

ABSTRACT

Ligand–receptor interactions are commonly assessed through equilibrium occupancy and pharmacodynamic measures that describe binding and saturation by means of bounded response curves. Thermodynamic approaches relate binding affinity to logarithmic concentration scaling, while probabilistic descriptions of occupancy arise from exponential relations. We introduce an exponential–logarithmic descriptor (ELD) that integrates ligand availability and thermodynamic binding propensity within a single quantity. The logarithmic component corresponds to a thermodynamic term derived from concentration-dependent free-energy relations, whereas the exponential component is represented through an inverse normalized concentration term corresponding to the reciprocal of the exponential occupancy factor emerging from Boltzmann-type binding formulations. We explored ELD behavior through numerical simulations spanning sub-affinity, transition and saturating concentration regimes under multiple affinity conditions and time-dependent exposure profiles. Compared with conventional occupancy curves, ELD retained a broader dynamic range and revealed asymmetric sensitivity across concentration scales, particularly at low exposure and near saturation, where bounded occupancy measures progressively compress variability. The resulting behavior reflects the coexistence of amplification and constraint processes within ligand–receptor dynamics. ELD may provide quantitative representation for biological systems in which exponential and logarithmic processes coexist across different scales. Potential applications include characterization of dose–response transitions, identification of subtherapeutic and saturating exposure states, comparison of compounds with different affinities, normalization across heterogeneous datasets and continuous tracking of pharmacodynamic regimes during time-dependent exposure.

**KEYWORDS:** nonlinearity; affinity; thermodynamics; scaling; sensitivity.

## INTRODUCTION

Ligand–receptor interactions are commonly described through equilibrium binding models relating ligand concentration to receptor occupancy by means of dissociation constants and saturating response curves (Eble 2018; Ng and Bauer 2024; Robertson et al. 2026). Rooted in classical biochemistry, these formulations provide measures of affinity and efficacy to characterize dose–response relationships in pharmacology (Yi et al. 2021; Chae et al. 2023; Ma et al. 2025; Shi 2026). Complementary thermodynamic approaches describe binding in terms of free-energy differences that vary logarithmically with concentration, thereby connecting molecular interactions to energetic constraints (Jandova et al. 2019; Alhadeff and Warshel 2020; Procacci 2021; Calderón et al. 2023; Azimi and Gallicchio 2024). In parallel, probabilistic descriptions of receptor occupancy arise from exponential relations derived from statistical mechanics (Jesudason et al. 2017; Madsen et al. 2019; Siafis et al. 2023). Despite their common theoretical basis, these perspectives are usually employed separately, with occupancy treated as the primary observable and thermodynamic quantities serving a supporting role. Standard representations tend to compress variability at both low and high concentration ranges, limiting sensitivity to regime transitions. Moreover, the bounded nature of occupancy measures can obscure differences between subtherapeutic and saturating conditions. This approach echoes both historical practice and the absence of a unified formulation combining these transformations into a single, operational metric.

We asked whether a single dimensionless descriptor based on exponential and logarithmic transformations could provide a coherent representation of ligand–receptor interactions across concentration regimes that are typically described separately by occupancy equations and thermodynamic relations. Our conceptual motivation is related to recent work by Ordzywolek (2026), who showed that a broad class of elementary mathematical functions can emerge from recursive combinations of exponential and logarithmic operations within a unified formal structure. Building on this idea, we examined whether these transformations could also provide a compact description of binding conditions already represented by established biochemical models. Our descriptor combines two complementary components derived from equilibrium binding relations. The logarithmic component $\ln([L]/K_d)$ originates from thermodynamic expressions for binding free energy and reflects the concentration-dependent energetic contribution to binding. The exponential-related component is represented by the inverse normalized concentration term $K_d/[L]$ corresponding to the reciprocal of the exponential occupancy factor arising in Boltzmann-type probabilistic descriptions of receptor binding. Together, these terms link ligand availability and thermodynamic binding propensity within a single quantity, without introducing additional parameters beyond those already present in equilibrium binding equations. Potential practical relevance may arise from the possibility of integrating ligand concentration and receptor affinity into a directly computable dimensionless measure that preserves clearer numerical separation between low-exposure, transition and saturation conditions than conventional occupancy curves. To assess our hypothesis, we performed numerical simulations across a broad range of normalized concentrations and compared the resulting responses with standard occupancy measures under different affinity and exposure conditions.

We will proceed as follows. First, we introduce the formal definition of the descriptor and its derivation from equilibrium relations. Then, we describe the simulation framework and comparative analysis. Finally, we report the results and examine their implications within standard pharmacological contexts.

## METHODS

We studied a simulated drug–receptor system in which ligand concentration, receptor affinity, receptor occupancy and an exponential–logarithmic descriptor were evaluated under static and time-dependent conditions. We used equilibrium binding equations, thermodynamic normalization, logarithmic concentration scaling, exponential inverse relations, numerical concentration sweeps, parameter-space mapping and simplified pharmacokinetic profiles. We aimed to identify an experimentally discriminable signature separating sub-affinity exposure, transition-zone binding and saturation, using variables expressed in real concentration units and compared with standard receptor occupancy.

**Model definition**. The simulated system consisted of a ligand $L$ interacting reversibly with a receptor $R$ to form a bound complex $RL$. The elementary binding scheme was written as

$$R + L \rightleftharpoons RL.$$

At equilibrium, the dissociation constant was defined as

$$K_d = \frac{[R][L]}{[RL]},$$

where $[L]$ denotes free ligand concentration, $[R]$ the concentration of unbound receptor and $[RL]$ the concentration of receptor–ligand complex. Total receptor concentration was defined as

$$R_T = [R] + [RL].$$

The fractional receptor occupancy was then obtained as

$$\theta = \frac{[RL]}{R_T}.$$

By substituting the equilibrium relation into the receptor mass balance, the standard occupancy equation followed:

$$\theta = \frac{[L]}{[L] + K_d}.$$

Occupancy was expressed as a percentage by multiplying by 100:

$$\theta_{\%} = 100\frac{[L]}{[L] + K_d}.$$

Ligand concentration was measured in nanomolar units and the dissociation constants used in the simulations were $K_d = 1,\ 10$ and $100$ nM. These values were not assigned to a specific drug class but were selected to represent high-, intermediate- and lower-affinity binding regimes within a common pharmacological range.

**Thermodynamic scaling**. To combine logarithmic and exponential terms without dimensional inconsistency, all concentration-dependent expressions were normalized by an affinity scale. The dimensionless exposure variable was defined as

$$z = \frac{[L]}{K_d}.$$

This transformation allowed ligand concentration and affinity to be compared in a unit-free form while retaining the original units in all plotted concentration axes. The thermodynamic relation between binding free energy and concentration ratio was expressed as

$$\Delta G = RT\ln\left(\frac{K_d}{[L]}\right),$$

where $R$ is the gas constant and $T$ the absolute temperature. Dividing by $RT$ gave the dimensionless free-energy ratio

$$g = \frac{\Delta G}{RT} = \ln\left(\frac{K_d}{[L]}\right) = -\ln\, z.$$

The corresponding Boltzmann-type expression for occupancy can be written as

$$\theta = \frac{1}{1 + \exp{(g)}}.$$

Substituting $g = -\ln\, z$ gives

$$\theta = \frac{1}{1 + \exp{(-\ln\, z)}} = \frac{1}{1 + z^{-1}} = \frac{z}{1 + z}.$$

Thus, the logarithmic and exponential transformations are both explicit in the thermodynamic representation, while the final occupancy equation takes the familiar saturating form.

**Descriptor construction**. The exponential–logarithmic descriptor was defined from the normalized concentration variable $z = [L]/K_d$. To represent the tendency toward unbound receptor availability, the inverse exposure term was used:

$$A(z) = \frac{1}{z} = \frac{K_d}{[L]}.$$

To represent the thermodynamic binding drive, the logarithmic term was written as

$$B(z) = \ln\, z = \ln\left(\frac{[L]}{K_d}\right).$$

The descriptor was then defined as

$$E(z) = A(z) - B(z) = \frac{1}{z} - \ln\, z.$$

In concentration variables, this becomes

$$E([L], K_d) = \frac{K_d}{[L]} - \ln\left(\frac{[L]}{K_d}\right).$$

The descriptor is dimensionless because both $K_d/[L]$ and $\ln\,([L]/K_d)$ are dimensionless. The first term decreases as ligand concentration exceeds the affinity scale, while the logarithmic term increases with ligand concentration. Therefore, high positive values occur at low exposure, values near unity occur around $[L] = K_d$ and negative values occur when ligand concentration exceeds affinity. The derivative was computed as

$$\frac{dE}{dz} = -\frac{1}{z^2} - \frac{1}{z},$$

showing that $E(z)$ decreases monotonically for $z > 0$.

**Concentration sweep**. Static simulations were performed over a logarithmically spaced ligand concentration range from 0.01 nM to 10,000 nM. The concentration vector was defined as

$$[L]_i = 10^{a_i},$$

with

$$a_i \in [-2,4],$$

sampled using 600 equally spaced points. For each ligand concentration and for each dissociation constant $K_d \in \{1,10,100\}$ nM, receptor occupancy and the exponential–logarithmic descriptor were computed as

$$\theta_i(K_d) = \frac{[L]_i}{[L]_i + K_d},$$

and

$$E_i(K_d) = \frac{K_d}{[L]_i} - \ln\left(\frac{[L]_i}{K_d}\right).$$

A logarithmic concentration axis was used because receptor binding spans orders of magnitude and because the relevant transition occurs near the ratio $[L]/K_d = 1$, rather than at a fixed absolute concentration. Three qualitative regions were identified for visual annotation: sub-affinity exposure, transition-zone binding and saturation. These regions were represented relative to the intermediate affinity condition $K_d = 10$ nM, using concentration intervals below, around and above $K_d$. No stochastic noise was added, because the purpose was to characterize the mathematical behavior of the descriptor under controlled deterministic conditions.

**Affinity landscape**. A two-dimensional concentration–affinity landscape was generated to evaluate how the descriptor behaved when ligand concentration and receptor affinity were varied simultaneously. Ligand concentration was sampled across

$$[L] \in [10^{-2}, 10^{4}] \text{ nM},$$

while dissociation constants were sampled across

$$K_d \in [10^{-1}, 10^{3}] \text{ nM}.$$

A rectangular grid was constructed by forming all pairs

$$([L]_i, K_{d,j}).$$

At each grid point, the normalized exposure was calculated as

$$z_{ij} = \frac{[L]_i}{K_{d,j}},$$

and the descriptor was computed as

$$E_{ij} = \frac{1}{z_{ij}} - \ln\, z_{ij}.$$

For visualization, extreme descriptor values were clipped to a finite plotting interval to prevent very large low-exposure values from dominating the color scale. This operation affected only graphical display and not the underlying equations. Occupancy was also computed at every grid point:

$$\theta_{ij} = \frac{[L]_i}{[L]_i + K_{d,j}}.$$

Contours corresponding to 10%, 50% and 90% occupancy were superimposed on the descriptor landscape. These contours were obtained from

$$\theta = \frac{z}{1+z},$$

which gives

$$z = \frac{\theta}{1-\theta}.$$

Thus, the plotted contour levels corresponded to $z = 1/9$, $z = 1$ and $z = 9$.

**Time simulation**. A simplified time-dependent exposure profile was generated to compare concentration, occupancy and descriptor dynamics after a single dose. The ligand concentration was modeled using a one-compartment oral-like difference of exponentials:

$$C(t) = S(e^{-k_e t} - e^{-k_a t}),$$

where $C(t)$ is plasma ligand concentration in nM, $k_a$ is the absorption rate constant, $k_e$ is the elimination rate constant and $S$is a scaling factor. Negative values before absorption were not permitted, so the computed profile was constrained by

$$C(t) = \max\,[0, S(e^{-k_e t} - e^{-k_a t})].$$

The profile was then normalized to a selected maximum concentration $C_{\max}$ through

$$C_{\text{scaled}}(t) = C_{\max} \frac{C(t)}{\max\limits_t\, C(t)}.$$

Time was sampled from 0 to 48 hours using 500 points. Two clearance scenarios were simulated. The standard-clearance condition used $C_{\max} = 100$ nM, $k_a = 1.1\ \text{h}^{-1}$ and $k_e = 0.18\ \text{h}^{-1}$. The reduced-clearance condition used $C_{\max} = 150$ nM, $k_a = 1.1\ \text{h}^{-1}$ and $k_e = 0.08\ \text{h}^{-1}$. The receptor affinity was fixed at $K_d = 10$ nM.

**Dynamic observables**. For each simulated concentration–time curve, receptor occupancy was calculated at every time point as

$$\theta(t) = \frac{C(t)}{C(t) + K_d}.$$

Occupancy in percent was then obtained as

$$\theta_{\%}(t) = 100\theta(t).$$

The time-dependent descriptor was computed from the same concentration profile using

$$z(t) = \frac{C(t)}{K_d},$$

and

$$E(t) = \frac{1}{z(t)} - \ln\, z(t).$$

Because $E(t)$ is undefined at exactly $C(t) = 0$, a small numerical lower bound was introduced only for computation:

$$C_\epsilon(t) = \max\,[C(t), \epsilon],$$

with

$$\epsilon = 10^{-6}\ \text{nM}.$$

The descriptor was therefore computed as

$$E(t) = \frac{K_d}{C_\epsilon(t)} - \ln\left(\frac{C_\epsilon(t)}{K_d}\right).$$

This avoided division by zero and logarithms of zero at the initial point. The dynamic comparison used identical $K_d$ values across the two clearance scenarios, isolating the effect of concentration–time evolution.

**Software tools**. All simulations and numerical analyses were performed using Python 3, with vectorized computations implemented through NumPy to ensure numerical stability and reproducibility across large parameter sweeps All equations were evaluated deterministically without stochastic components and no random sampling procedures were used.

## RESULTS

We report the behavior of our exponential–logarithmic descriptor across static and time-dependent ligand–receptor conditions, spanning multiple affinity scales and concentration regimes. Simulations examined concentration-dependent occupancy, normalized descriptor dynamics and pharmacokinetic exposure profiles under different clearance conditions.

**Static regimes**. Across ligand concentrations ranging from 0.01to 10,000 nM, receptor occupancy showed the expected sigmoidal transition from low binding to saturation for all dissociation constants examined ($K_d = 1$, 10and 100 nM). The left panel of Figure 1 shows classical receptor occupancy as a function of concentration, illustrating progressive saturation and reduced sensitivity at both low- and high-concentration extremes. For $K_d = 10$ nM, occupancy increased from 0.99%at 0.1 nMto 50%at 10 nM, reaching 90.9%at 100 nM. In contrast, the exponential–logarithmic descriptor exhibited a substantially broader numerical range. The upper-right panel of Figure 1 reports the descriptor as a function of normalized concentration, showing an extended dynamic range and asymmetric transitions across concentration regimes. Under the same affinity condition, descriptor values decreased from approximately 104.6at 0.1 nMto 12.3at 1 nM, 1.0at 10 nM, and −2.2at 100 nM. Therefore, our descriptor crossed from large positive values to negative values over the same concentration interval in which occupancy approached saturation. In the lower-right panel, the concentration–affinity landscape mapped descriptor values across ligand concentration and dissociation constants, with occupancy contours at 10%, 50% and 90%superimposed on the surface.
The integrated representation in Figure 1 shows that occupancy compresses variability into bounded asymptotic regions, whereas our descriptor preserves continuous separation between sub-affinity, transition and saturation conditions across multiple affinity scales.

**Dynamic exposure**. Time-dependent simulations over 48 hours showed distinct concentration trajectories for standard and reduced clearance conditions. The upper-left panel of Figure 2 displays concentration–time curves for both elimination profiles together with the reference affinity constant. Peak concentrations reached approximately 100 nMin the standard-elimination profile and 150 nMin the reduced-clearance condition. Occupancy rapidly exceeded 90% in both cases and remained near saturation during the early phase of exposure despite differences in concentration magnitude. The lower-left panel of Figure 2 shows the corresponding occupancy trajectories, illustrating limited separation between clearance conditions during the high-exposure phase. Under standard elimination, occupancy declined below 50% after approximately 13 hours, whereas reduced clearance delayed this transition to approximately 24 hours. The right panel of Figure 2 illustrates the temporal evolution of our exponential–logarithmic descriptor computed from instantaneous concentration relative to affinity. During the early high-exposure phase, descriptor values were negative, ranging between approximately −2 and −3. As concentration approached the affinity scale, descriptor values increased toward 1, subsequently becoming positive during sub-affinity exposure. Reduced clearance produced a prolonged interval of negative descriptor values and delayed return toward positive values compared with standard elimination. Unlike occupancy, which remained compressed near its upper bound during saturation, our descriptor continuously reflected concentration changes relative to affinity throughout the simulated trajectories.

Overall, our exponential–logarithmic descriptor and conventional occupancy measures could provide distinct but complementary representations of ligand–receptor interactions. Occupancy retains its characteristic bounded saturating behavior, whereas our descriptor preserves broader numerical variation across low-exposure, transition and saturation regimes. This suggests that the observed dynamics arise from the combination of inverse concentration scaling and logarithmic thermodynamic transformation within a single dimensionless quantity directly linked to measurable biochemical variables.

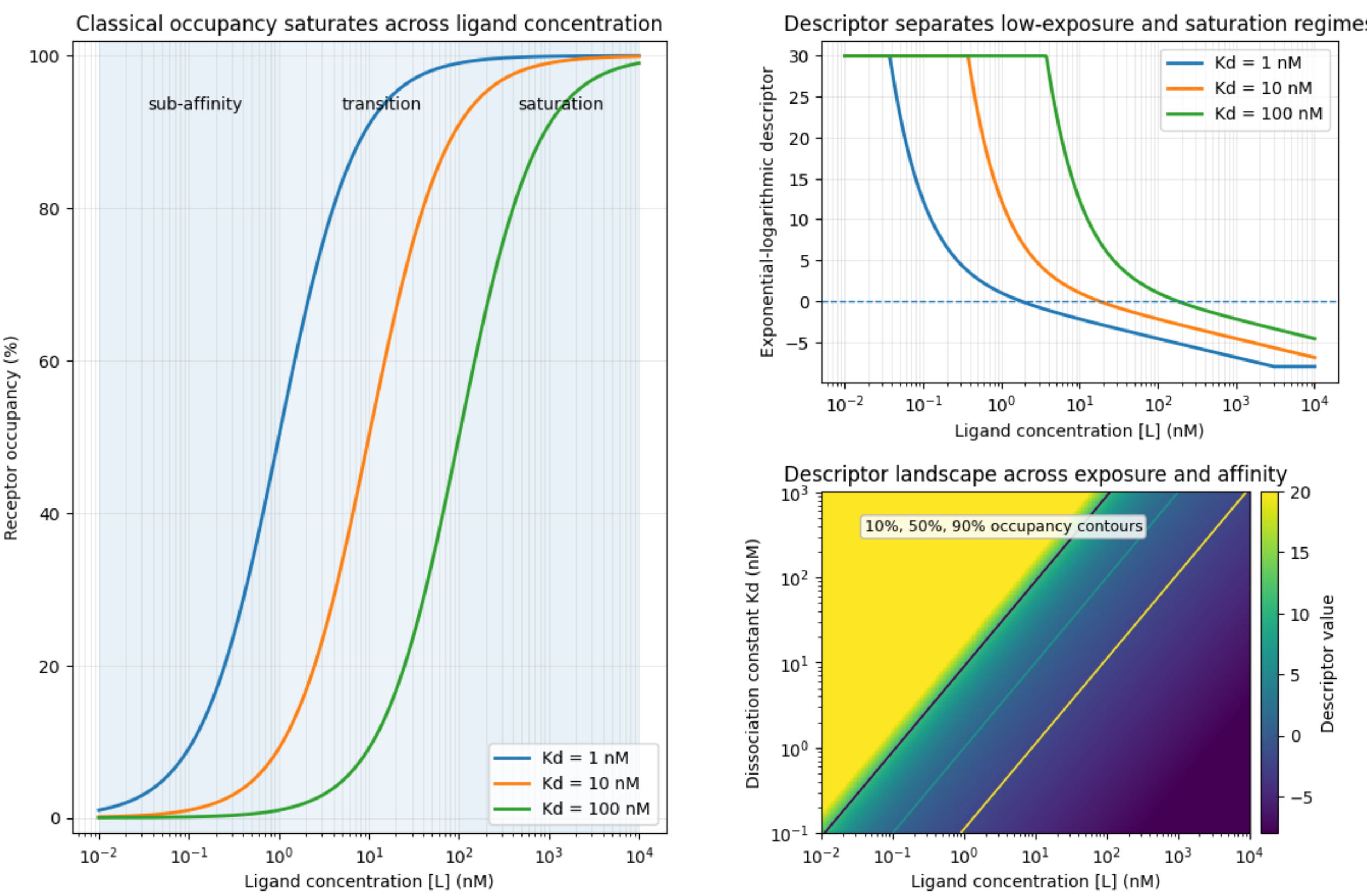

**Figure 1**. Representation of ligand–receptor binding, saturation and exponential–logarithmic descriptor across concentration and affinity scales. Simulated ligand–receptor interactions are displayed over ligand concentrations ranging from 0.01 to 10,000 nM for dissociation constants of 1, 10 and 100 nM.
The left panel shows classical receptor occupancy as a function of concentration, illustrating the characteristic sigmoidal transition from low binding to saturation. Note the reduced sensitivity at both extremes.
The upper-right panel reports the exponential–logarithmic descriptor computed from normalized concentration, showing extended dynamic range and asymmetric behavior across regimes.
The lower-right panel presents a two-dimensional concentration–affinity landscape, where the descriptor is mapped across both ligand concentration and dissociation constant, with superimposed occupancy contours at 10%, 50% and 90%.
In sum, occupancy compresses information at high and low concentrations, while our descriptor preserves separation between sub-affinity, transition and saturation regimes across varying affinities.

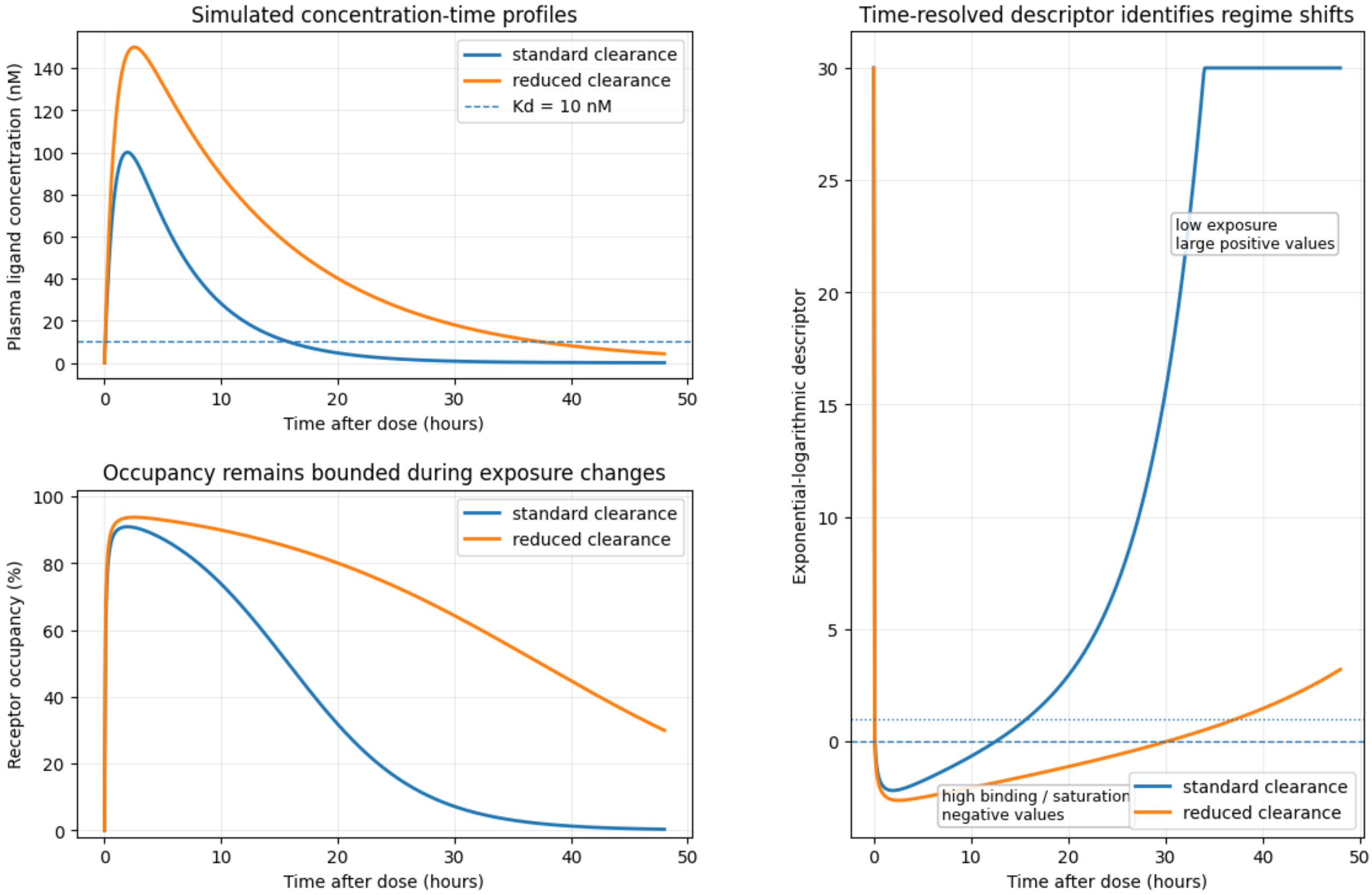

**Figure 2.** Time-resolved dynamics of ligand exposure, receptor occupancy and exponential–logarithmic descriptor under different clearance conditions. Simulated pharmacokinetic profiles following a single dose are shown over 48 hours, with plasma ligand concentration expressed in nM for two clearance scenarios: standard elimination and reduced clearance.
The upper-left panel displays concentration–time curves, including the reference dissociation constant (Kd = 10 nM).
The lower-left panel shows corresponding receptor occupancy, illustrating rapid approach to high occupancy followed by gradual decline, with limited discrimination between clearance conditions at high exposure.
The right panel presents the time evolution of the exponential–logarithmic descriptor, computed from instantaneous concentration relative to Kd.
In sum, our descriptor emphasizes differences in exposure dynamics that are less apparent from occupancy alone. It distinguishes early saturation, intermediate transition and late low-exposure phases, revealing delayed recovery toward baseline under reduced clearance.

## CONCLUSIONS

We asked whether a single quantitative descriptor combining exponential and logarithmic transformations could provide a coherent representation of ligand–receptor interactions across concentration regimes typically described separately by occupancy and thermodynamic relations. Our simulations compared classical receptor occupancy derived from equilibrium binding equations with a dimensionless exponential–logarithmic descriptor evaluated over wide concentration ranges and under time-dependent exposure conditions. We found that occupancy retained its expected sigmoidal and bounded behavior, approaching saturation as ligand concentration exceeded the dissociation constant, while our descriptor varied over a broader numerical domain and decreased monotonically with normalized concentration. In static analyses, our descriptor preserved separation between sub-affinity, transition and saturation regimes even where occupancy approached asymptotic limits. In turn, in time-resolved simulations, our descriptor reflected changes in exposure relative to affinity more continuously, distinguishing phases of high binding, intermediate transition and low exposure as concentration evolved.
Our descriptor combines an inverse concentration term with a logarithmic transformation derived from thermodynamic relations, capturing the interplay between ligand availability and binding propensity within a single quantity. Directly linked to measurable variables, it preserves sensitivity across concentration regimes that become progressively compressed in conventional occupancy representations.

Our framework differs from commonly used approaches such as Michaelis–Menten kinetics, Hill-type dose–response models and standard receptor occupancy formulations, which rely on bounded nonlinear curves compressing variability at extreme concentration ranges (Schindler 2017; Seibert and Tracy 2021; Srinivasan 2022; Schindler 2022). Unlike these techniques which primarily quantify fractional engagement, our formulation produces an unbounded, continuous measure derived directly from normalized concentration ratios, preserving sensitivity across multiple scales. Compared with thermodynamic free-energy representations based on Gibbs free energy (Clark et al. 2023; Jia et al. 2023; Azimi and Gallicchio 2024), our descriptor combines inverse and logarithmic transformations into a single expression that is directly computable from measurable quantities.

Our study has limitations. The exponential–logarithmic descriptor $E = 1/z - \ln z$ is built ad hoc and is not derived from a conservation law, variational principle or established physical observable. The use of thermodynamic relations is incomplete: although $\Delta G = RT\ln(K_d/[L])$ is valid, its incorporation into the descriptor does not follow directly from equilibrium thermodynamics. Our pharmacokinetic representation relies on a simplified difference-of-exponentials model that omits distribution volumes, binding compartments and nonlinear clearance. No statistical inference was performed and regime separation is described qualitatively. Generalizability is limited to systems approximated by equilibrium binding with well-defined $K_d$ and measurable concentrations. The relation to single-operator formulations is conceptual and not empirically demonstrated.

Testable hypotheses can be drawn. First, in equilibrium binding assays, values of $E$ should align with occupancy regimes in a concentration-invariant manner when plotted against the normalized ratio $z = [L]/K_d$. Quantitatively, concentrations satisfying $z \approx 1$ should consistently yield $E \approx 1$, while $z = 0.1$ and $z = 10$ should produce $E \approx 12.3$ and $E \approx -2.3$, respectively, independent of the absolute scale of $K_d$. This predicts collapse of data from ligands with different affinities onto a single curve when expressed in terms of $z$.
Second, in time-resolved pharmacokinetic–pharmacodynamic experiments, the temporal crossing of $E = 1$ should occur when instantaneous concentration equals the dissociation constant. Measuring concentration–time profiles and receptor occupancy simultaneously should show that the transition from $E > 1$ to $E < 1$ coincides with occupancy near 50%, providing a direct experimental signature which links descriptor dynamics to binding equilibrium.
Third, under conditions of impaired clearance, the duration of negative $E$ values should scale with the elimination rate constant. Specifically, reducing $k_e$ in pharmacokinetic models or clinical measurements should prolong the time interval $t$ such that $E(t) < 0$, with a predictable shift proportional to the inverse of $k_e$.
Future research may address whether a mechanistic derivation of the descriptor can be achieved, extend the formulation to cooperative or multivalent binding systems and test its behavior in nonequilibrium conditions where binding kinetics and transport processes interact.

Practical applications may arise in contexts where compact descriptors are required to compare heterogeneous datasets without reparameterization. Our descriptor can be computed directly from routinely measured concentrations and reported affinity values, enabling rapid normalization across compounds or experimental conditions. It may support data integration workflows in which measurements obtained under different assay formats or concentration ranges need to be aligned onto a common scale. In longitudinal monitoring, it could help in tracking deviations from reference states using a single scalar variable, facilitating visualization of trajectories without relying on multiple parallel metrics. It may also be incorporated into algorithmic pipelines for model calibration, where dimensional reduction of input variables is required before optimization or classification procedures.

In conclusion, we introduced a dimensionless descriptor combining inverse and logarithmic transformations of normalized concentration and examined its theoretical behavior under simulated ligand–receptor binding conditions. The observed dynamics indicate that our formulation preserves sensitivity across concentration regimes, while remaining directly linked to measurable biochemical variables. Rather than replacing established occupancy or thermodynamic measures, our descriptor could provide a quantitative representation that integrates information distributed across various analytical descriptions.

## DECLARATIONS

**Ethics approval and consent to participate.** This research does not contain any studies with human participants or animals performed by the Author.
**Consent for publication.** The Author transfers all copyright ownership, in the event the work is published. The undersigned author warrants that the article is original, does not infringe on any copyright or other proprietary right of any third part, is not under consideration by another journal and has not been previously published.
**Availability of data and materials.** All data and materials generated or analyzed during this study are included in the manuscript. The Author had full access to all the data in the study and took responsibility for the integrity of the data and the accuracy of the data analysis.

**Disclaimer**. The views expressed are those of the author and do not necessarily reflect those of the affiliated institutions.
**Competing interests.** The Author does not have any known or potential conflict of interest including any financial, personal or other relationships with other people or organizations within three years of beginning the submitted work that could inappropriately influence or be perceived to influence their work.
**Funding.** This research did not receive any specific grant from funding agencies in the public, commercial or not-for-profit sectors.
**Acknowledgements:** none.
**Authors' contributions.** The Author performed: study concept and design, acquisition of data, analysis and interpretation of data, drafting of the manuscript, critical revision of the manuscript for important intellectual content, statistical analysis, obtained funding, administrative, technical and material support, study supervision.
**Declaration of generative AI and AI-assisted technologies in the writing process.** During the preparation of this work, the author used ChatGPT 5.3 to assist with data analysis and manuscript drafting and to improve spelling, grammar and general editing. After using this tool, the author reviewed and edited the content as needed, taking full responsibility for the content of the publication.